\documentclass[9pt,twocolumn,twoside]{osajnl}
\usepackage{mathrsfs}
\usepackage{color}

\journal{ol} 

\setboolean{shortarticle}{true}

\title{Simple phase noise measurement scheme for cavity-stabilized laser systems}

\author[1, *]{Fabian Schmid}
\author[1]{Johannes Weitenberg}
\author[1, 2]{Theodor W. Hänsch}
\author[1, 2]{Thomas Udem}
\author[1]{Akira Ozawa}

\affil[1]{Max-Planck-Institut für Quantenoptik, 85748 Garching, Germany}
\affil[2]{Fakultät für Physik, Ludwig-Maximilians-Universität München, 80799 München, Germany}
\affil[*]{Corresponding author: fabian.schmid@mpq.mpg.de}

\dates{}

 \doi{\url{https://doi.org/10.1364/OL.44.002709}}

 \renewcommand{\copyrightstatement}{\, \textcopyright\, 2019{} Optical Society of America. One print or electronic copy may be made for personal use only. Systematic reproduction and distribution, duplication of any material in this paper for a fee or for commercial purposes, or modifications of the content of this paper are prohibited.}

\begin{abstract}
    We describe a simple method for measuring the residual fast phase noise of a cavity-stabilized laser using the cavity as a reference. The method is based on generating a beat note between the laser output and the strongly filtered light transmitted through the cavity. The beat note can be directly analyzed without requiring further calibration of system parameters. We apply the method to measure the residual phase noise of an external-cavity diode laser (ECDL) locked to a reference cavity and compare the results with an analysis of the in-loop error signal of the feedback system.
\end{abstract}

\setboolean{displaycopyright}{true}

\begin{document}

\maketitle
\begin{figure}
    \includegraphics[width=\linewidth]{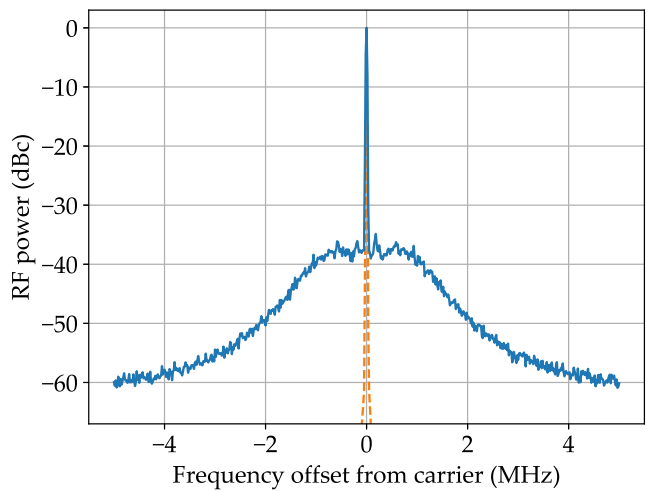}
    \caption{Typical beat note between two ECDLs which are stabilized to independent reference cavities and have subhertz linewidths (solid blue trace) recorded with a resolution bandwidth of 20~kHz. The narrow carrier (resolution bandwidth limited) and broad phase-noise pedestal are clearly visible. The orange dashed line shows the expected beat note after filtering the light of both lasers through a cavity with a FWHM linewidth of 6.55~kHz.}
    \label{fig:beat_note}
\end{figure}

Lasers stabilized to high-finesse cavities have become an important tool in a wide range of fields, such as high-resolution laser spectroscopy \cite{Kolachevsky2011}, gravitational wave detection \cite{Kwee2012}, optical clocks \cite{Bloom2014, Huntemann2016}, and low-noise microwave generation \cite{Fortier2011, Xie2016}. A common figure of merit for a stabilized laser is its full width at half maximum (FWHM) linewidth. Two limiting regions of phase/frequency noise may be distinguished: Low Fourier frequency with a large phase range (slow frequency noise) and high Fourier frequency with a small phase range (fast frequency noise). The two spectral regions are separated by the point where the frequency noise power spectral density $S_\nu^I(f)$ crosses the so-called $\beta$-separation line $b f$ with a dimensionless constant $b \sim 1$ \cite{Domenico2010}. Slow frequency noise often gives rise to a Gaussian line shape while fast frequency noise, like shot noise, gives rise to a Lorentzian line shape \cite{Elliott1982, Domenico2010}. The linewidth of a laser is determined by slow frequency noise, while fast frequency noise only contributes to the wings of the spectrum and has no significant effect on the linewidth \cite{Domenico2010}. Therefore, FWHM linewidth is not always a good measure for the spectral purity of a laser. At low Fourier frequencies, the feedback loop that stabilizes the laser can have very large gain, tightly locking the laser to the cavity resonance. The linewidth is therefore often determined by the stability of the reference cavity. By employing cryogenically cooled single-crystal silicon cavities, linewidths below 10~mHz have been achieved \cite{Matei2017}. The situation is different for noise at high Fourier frequencies. Feedback loop stability requires that the feedback gain drops below 0~dB before a phase delay of 180$^\circ$ is reached. Depending on the loop layout and the transducers used, this point is typically reached for frequencies in the range of a few tens of kHz to a few MHz. Therefore, the intrinsic noise of the laser can only partially be suppressed. Consequently, even for ultrastable lasers with subhertz linewidths, a significant fraction of the overall power can be contained in a broad phase-noise ``pedestal'' surrounding the narrow carrier. This is illustrated in Fig.~\ref{fig:beat_note} which shows a typical beat note between two ECDLs which are stabilized to independent reference cavities and have subhertz linewidths. Since the two laser systems are very similar in construction and have uncorrelated outputs, the noise power in the beat note spectrum is approximately twice the noise power of each individual laser. The relative power contained in the carrier of a signal with phase noise is $\exp(- \phi_\text{rms}^2)$, where $\phi_\text{rms}$ is the rms integrated phase noise \cite{Telle1996}. With a good feedback loop, an integrated phase noise on the order of 100~mrad can be achieved even with a very simple ECDL \cite{Kolachevsky2011}. Therefore, at the fundamental wavelength, the power lost from the carrier is typically a few percent at most. The situation changes when harmonics of the laser output are generated. The rms phase noise at the $q$th harmonic is given by $q \phi_\text{rms}$, and the power contained in the upconverted carrier is $\exp(- q^2 \phi_\text{rms}^2)$. Once the argument of the exponential function surpasses a threshold of (1~rad)$^2$, the carrier starts to collapse \cite{Telle1996}. High harmonic generation of frequency combs has been used to extend high resolution spectroscopy to the extreme ultraviolet (XUV) \cite{Kandula2010, Cingoez2012, Ozawa2013}. So far this technique was only applied to broad transitions with MHz linewidths where the noise pedestal of the laser spectrum can contribute to the excitation rate. It is expected that much higher measurement accuracies can be achieved by addressing narrower transitions such as the 1S--2S transition in He$^+$ \cite{Herrmann2009}. In this case, only the power in the carrier can effectively contribute to driving the excitation. While high harmonic generation has been shown not to cause additional linewidth broadening \cite{Benko2014}, the amount of phase noise in the driving laser limits the harmonic order that can be achieved before the carrier collapses \cite{Corsi2017}, making the generated light unusable for this application. In some applications, the presence of the noise sidebands can also directly affect the system performance. In photonic microwave generation the frequency of a reference laser is subdivided to the radio frequency domain, and the phase noise of the laser ultimately sets the limit for the spectral purity of the generated RF signal \cite{Xie2016}. High resolution spectroscopy of broad transitions requires finding the line center within a small fraction of the transition linewidth \cite{Beyer2017}. Since the spectrum of the laser is folded into the observed line, a possible asymmetry in the noise pedestal due to correlated amplitude and phase noise can lead to a systematic error in this determination. In such noise-sensitive applications, evaluating the linewidth does not give sufficient information whether the laser is low-noise enough. It is important to characterize the phase noise spectrum over a wide Fourier frequency range.

The standard method for a complete phase noise characterization of a laser system is to compare it with a second independent laser system with equal or better noise performance. Information about the phase noise spectrum can also be obtained using a delayed self-heterodyne interferometer with a fiber delay line \cite{Tsuchida2011}. In this Letter we show how the residual fast phase noise of a cavity-stabilized laser can be characterized by generating a beat note between the laser output and the light transmitted through the cavity. Since the cavity averages the amplitude and phase fluctuations of the incident light over the photon lifetime in the cavity, faster noise components are strongly attenuated in the light circulating in the cavity and therefore also in the cavity transmission. By comparing the laser output to the cavity transmission, the noise with Fourier frequencies above the cavity linewidth can be evaluated.

\begin{figure}
    \includegraphics[width=\linewidth]{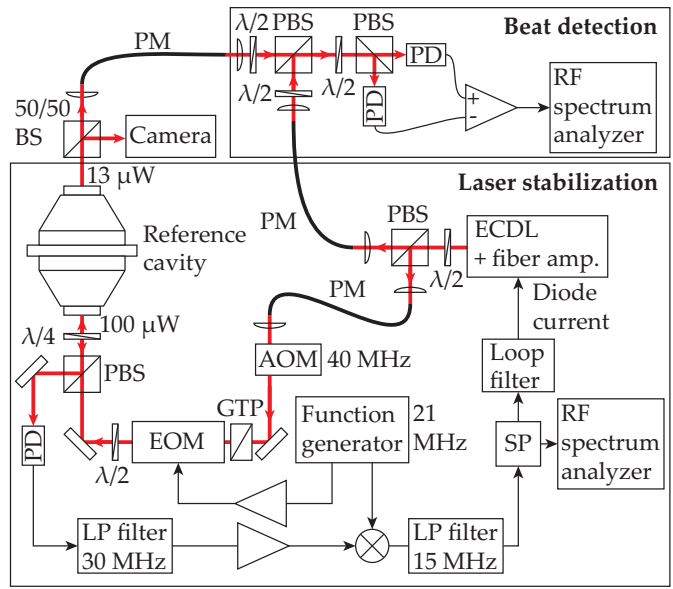}
    \caption{Experimental setup. The lower box shows the hardware for laser stabilization, while the upper box shows the additional components required for the method described here. The beat note is generated on a balanced photodetector \cite{Carleton1968}. PBS, polarizing beam splitter. AOM, acousto-optic modulator. GTP, Glan-Taylor polarizer. EOM, electro-optic modulator. BS, non-polarizing beam splitter. PD, photodiode. LP filter, low-pass filter. PM, polarization-maintaining fiber. SP, RF splitter.}
    \label{fig:setup}
\end{figure}

Fig.~\ref{fig:setup} shows a schematic of the experimental setup. The laser system consists of a home-built interference-filter stabilized ECDL \cite{Baillard2006} operating at 1033~nm and an Yb-doped fiber amplifier. Part of the light is sent through a polarization-maintaining fiber to the reference cavity setup which is placed on an active vibration isolation table. The frequency of the light is shifted by 40~MHz using an acousto-optic modulator (AOM) such that the light from the cavity and the laser output can eventually produce a beat note at this frequency. The AOM can also be used for fiber noise cancellation \cite{Ma1994} with a setup similar to \cite{Alnis2008}. Since the measurements reported here are made at frequencies above the acoustic range, fiber noise cancellation was omitted and the AOM is driven at a fixed frequency. An electro-optic modulator (EOM) is used to generate the sidebands required for locking the laser to the cavity using the Pound-Drever-Hall (PDH) technique \cite{Drever1983}. The modulation frequency is 21.32~MHz, and the modulation index is $\beta = 0.95$. The cavity consists of a 77.5~mm long spacer made from ultra-low expansion (ULE) glass and optically contacted dielectric mirrors with ULE substrates. A photon lifetime of $\tau_c = 24.3~\mu$s (relative drop of transmitted power to $1/e$) corresponding to a FWHM linewidth of 6.55~kHz was measured using the ring-down method. With a free spectral range of 1.93~GHz this results in a finesse of 295,000. The cavity is temperature stabilized and located in a vacuum chamber in a setup similar to FP2 in \cite{Alnis2008}. We send about 100~$\mu$W of light to the cavity, coupling in about 60\% of the carrier power. The light reflected from the cavity is sent onto an InGaAs photodiode (Hamamatsu G8376-008A) followed by a transimpedance amplifier with a gain of 10~kV/A. The signal is low-pass filtered at 30~MHz in order to suppress the influence of the second-order sidebands at twice the modulation frequency. After further amplification and demodulation in an RF mixer, the resulting PDH error signal is sent to a commercial loop filter (Toptica FALC 110) which gives feedback to the laser diode current of the ECDL.

In transmission a Fabry-P\'erot cavity acts as a first-order low-pass filter for amplitude and phase noise with a corner frequency corresponding to its half width at half maximum (HWHM) linewidth $\Delta f_\text{HWHM} = 1/(4 \pi \tau_c)$. Therefore, the noise components falling outside the cavity linewidth are strongly attenuated with a slope of -20~dB/decade \cite{Hald2005}. The expected filtering effect of our reference cavity is visualized in Fig.~\ref{fig:beat_note}. The transmission of a high-finesse reference cavity can be used as a low-noise laser source \cite{Nazarova2008}. However, the available power is often very small since the power sent to the reference cavity has to be limited in order to avoid drifts of the resonance frequency \cite{Alnis2008}. In our setup 13~$\mu$W are transmitted through the cavity. Half of the light is sent onto a camera in order to monitor the spatial mode inside the cavity. The other half is coupled into a single-mode fiber and sent to a beat note detection setup (see Fig.~\ref{fig:setup}). There it is overlapped with about 1~mW of light from the cavity-stabilized ECDL, and a beat note is generated on a balanced photodetector \cite{Carleton1968}. Compared to using a single detector, this improves the achievable signal-to-noise ratio and suppresses possible amplitude noise in the laser output. The beat note is analyzed by an RF spectrum analyzer (Agilent E4440A). The single sideband phase noise $\mathscr{L}(f)$ is defined as the noise power spectral density relative to the carrier power at frequency offset $f$ from the carrier. It is important to note that the envelope detection and logarithmic processing traditionally used in spectrum analyzers lead to systematic errors when measuring noise powers \cite{Keysight2017}. Many modern spectrum analyzers provide alternative detector options that avoid these issues. We use the integrated phase noise measurement function of our spectrum analyzer that automatically selects the correct detector, performs a logarithmic sweep of the frequency offset, and normalizes the measured traces according to the resolution bandwidth and carrier power. 

\begin{figure}[h]
    \includegraphics[width=\linewidth]{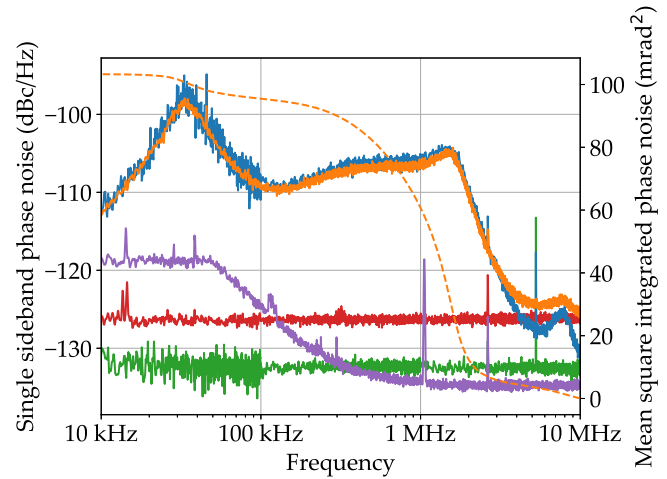}
    \caption{Single sideband phase noise $\mathscr{L}(f)$ between the cavity-stabilized laser and its reference cavity measured using the beat note with the cavity transmission (orange trace) and extracted from the in-loop error signal (blue trace). The noise floor originating from the beat note photodetection (red trace) is obtained by repeating the measurement with the cavity transmission light blocked. The phase noise of the reference oscillator inside the spectrum analyzer and of the signal generator that drives the frequency-shifting AOM is measured by directly connecting the signal generator to the spectrum analyzer (purple trace). The dashed orange curve shows the mean square integrated phase noise calculated from the beat note of the laser with the cavity transmission. An upper limit for the noise level of the PDH detection (green trace) is obtained by repeating the measurement with the feedback loop turned off and the laser detuned from the cavity resonance, such that the entire incident power is reflected onto the PDH photodiode. The spurious peak close to 5~MHz is an intermodulation product generated in the PDH detection.}
    \label{fig:phase_noise_spectrum}
\end{figure}

The orange trace in Fig.~\ref{fig:phase_noise_spectrum} shows the resulting phase noise spectrum for Fourier frequencies between 10~kHz and 10~MHz. In this frequency range, the phase noise in the light transmitted through the cavity is suppressed by more than an order of magnitude and therefore does not affect the measurement. The phase noise of the laser reaches a maximum value of -98~dBc/Hz at 30~kHz which we attribute to technical noise on the ECDL that is not completely suppressed by the feedback loop. The slight bump at 1.5~MHz is caused by the feedback loop whose phase margin reaches zero at this frequency. The phase noise of the reference oscillator inside the spectrum analyzer and of the signal generator that drives the frequency-shifting AOM (purple trace) contributes to the measured noise levels. It contains some spurious signals, but remains well below the phase noise of the laser in the entire Fourier frequency range. The photodetector noise (red trace) becomes comparable to the phase noise of the laser for Fourier frequencies above 3~MHz which sets the sensitivity limit in our particular setup. The integrated phase noise starting from 10~MHz is plotted as a dashed orange line. It shows that most of the phase noise contribution comes from the Fourier frequency range between 2 MHz and a few hundred kHz where the limited gain of the feedback loop cannot fully suppress the intrinsic noise of the laser. The total integrated phase noise from 10~MHz to 10~kHz is $\phi_\text{rms} = 10.2$~mrad. In an upcoming experiment we plan to excite the 1S--2S two-photon transition in He$^+$ ions \cite{Herrmann2009} at 60.8~nm using the 17th harmonic of a frequency comb centered at 1033~nm. Due to the two-photon absorption, this effectively corresponds to generating the 34th harmonic of the infrared light. Based on our results we expect that if the stability of our CW laser can be faithfully transferred to a frequency comb, 89\% of the power remain within 10~kHz around the carrier after the upconversion.

Another method for measuring the noise performance of a laser relative to its reference is to analyze the in-loop error signal of the lock. We use a wideband RF splitter (minicircuits ZFRSC-2050+) in order to send the same error signal to the loop filter and to a spectrum analyzer while maintaining impedance matching between the components. The spectrum analyzer measures the RF power $p(f)$ of the signal on its input contained within the chosen resolution bandwidth at frequency $f$. The single-sided voltage power spectral density of the signal is obtained by normalizing to a 1~Hz bandwidth and multiplying with the system impedance:
\begin{equation}
    S^\text{I}(f) = Z_0 p(f)/\text{RBW},
    \label{eqn:voltage_psd}
\end{equation}
where $Z_0 =$ 50~$\Omega$ is the system impedance, and $\text{RBW}$ is the resolution bandwidth of the spectrum analyzer.

In order to convert voltage fluctuations of the error signal to frequency fluctuations of the laser, the slope of the PDH error signal has to be determined. Since the free-running linewidth of our laser is much broader than the cavity line, we cannot directly observe the slope by sweeping the laser over the resonance. Instead, we lock the laser to the cavity and slowly modulate the setpoint of the loop filter at a frequency of about 100~Hz while recording the light intensity transmitted through the cavity. The drop in power transmitted through the cavity can be easily converted to a frequency offset using the known cavity linewidth \cite{Gatti2015}. In this way, we measure a slope of the PDH error signal at DC of $k_0 = 2.30 \times 10^{-4}$~V/Hz. At Fourier frequencies above the cavity linewidth, the field stored inside the cavity can no longer follow the fluctuations of the incident field. Therefore, the slope of the PDH discriminator has the frequency-dependent form \cite{Nagourney2014}
\begin{equation}
    k(f) = \frac{k_0}{\sqrt{1+4 (f/\Delta f_\text{FWHM})^2}},
    \label{eqn:pdh_tf}
\end{equation}
where $f$ is the frequency of the signal, and $\Delta f_\text{FWHM}$ is the FWHM linewidth of the cavity.

The phase noise of the laser can be obtained using the transfer function:
\begin{equation}
    \mathscr{L}(f) = S_\phi^\text{I}(f)/2 = S_\nu^\text{I}(f)/(2 f^2) = S^\text{I}(f)/(2 f^2 k^2(f)),
    \label{eqn:phase_psd}
\end{equation}
where $S_\nu^\text{I}(f)$ and $S_\phi^\text{I}(f)$ are the single-sided frequency and phase noise power spectral densities.

The result of the in-loop characterization is shown as the blue trace in Fig.~\ref{fig:phase_noise_spectrum}. It is in good agreement with the measurement using the cavity transmission. One complication when analyzing the in-loop error signal is that for sufficiently large gain the feedback loop can suppress the error below the error signal detection noise limit, thereby writing the noise onto the system output. In this case, the analysis gives an underestimation of the system noise. In our measurement the detection noise limit (green trace) stays below the measured in-loop error signal in the analyzed Fourier frequency range which ensures that the feedback loop does not add significant noise from our detection setup to the laser system.

While analyzing the in-loop error signal does not require setting up additional optics, the method proposed here has the advantage that it produces an RF beat note that can be directly analyzed without the need for an error-prone calibration procedure. It may appear that the in-loop characterization has the additional advantage that the phase noise at Fourier frequencies within the cavity linewidth can be evaluated. However, the phase noise of a stabilized laser in this frequency range is often dominated by slow drifts and fluctuations of the cavity which do not appear on the measurement using the cavity as a reference.  An out-of-loop comparison with a second independent setup would be necessary to evaluate such contributions.

In conclusion, we have shown a simple method for characterization of the residual fast phase noise of a cavity-stabilized laser at Fourier frequencies above the cavity linewidth. The noise in this frequency range is typically given by the inherent noise level of the laser and by the feedback bandwidth, rather then by the quality of the reference cavity. It is therefore a useful tool for evaluating the performance of the laser itself and for optimization of the feedback loop. Since reference cavity setups often already contain frequency-shifting AOMs, and the transmitted light is usually accessible, the method can be easily implemented in existing setups.

\section*{Funding}
T.W.H. acknowledges support from the Max Planck Foundation. European Research Council (ERC) (742247).
\section*{Acknowledgments}
We thank Lothar Maisenbacher for help in the experiment.

\bibliography{phase_noise}




\end{document}